%% file: main.tex
\documentclass[conference]{IEEEtran}  
\IEEEoverridecommandlockouts                              

\usepackage{amsmath} 

\usepackage{upgreek}
\usepackage[dvipsnames,table]{xcolor}

\usepackage{pgf}

\usepackage{subfig}
\usepackage{float}
\usepackage[export]{adjustbox}

\usepackage{algorithm}
\usepackage{algorithmicx,algpseudocode}
\usepackage[normalem]{ulem}
\graphicspath{{./figures/}}
\usepackage{color}
\usepackage{soul}
\usepackage{cite}
\usepackage{amsmath,amssymb,amsfonts}

\usepackage{graphicx}
\usepackage{textcomp}
\usepackage{xcolor}
\def\BibTeX{{\rm B\kern-.05em{\sc i\kern-.025em b}\kern-.08em
    T\kern-.1667em\lower.7ex\hbox{E}\kern-.125emX}}

\usepackage[a4paper,left=0.511811in,right=0.511811in,top=0.75in,bottom=1.7in]{geometry}
\usepackage[T1]{fontenc}

\usepackage[style=super,nopostdot, nonumberlist,acronym,nogroupskip]{glossaries}
\makeglossaries

\title{
On the impact of numerology in NR V2X Mode 2 with sensing and no-sensing resource selection
}

\author{
\IEEEauthorblockN{
Zoraze Ali, Sandra Lag\'en, Lorenza Giupponi
	}
\IEEEauthorblockA{
Centre Tecnol\`ogic de Telecomunicacions de Catalunya (CTTC/CERCA), Barcelona, Spain \\  emails:\{zoraze.ali, sandra.lagen, lorenza.giupponi\}@cttc.es
}
}

\begin{document}

\maketitle
\input{glossary}

\begin{abstract}
\input{abstract.tex}
\end{abstract}



\maketitle

\section{Introduction}
\label{sec:introduction}
\input{introduction.tex}

\section{NR V2X Mode 2}
\label{sec:2}
\input{sec-2.tex}

\section{Numerologies' Impact on NR V2X Mode 2}
\label{sec:3}
\input{sec-3.tex}

\section{Simulation Results}
\label{sec:4}
\input{sec-4.tex}

\section{Conclusions}
\label{sec:conclusions}
\input{conclusions.tex}

\section*{Acknowledgements}\small
This work was partially funded by Spanish MINECO grant TEC2017-88373-R (5G-REFINE), Generalitat de Catalunya grant 2017 SGR 1195, and the National Institute of Standards and Technology (NIST) from the U.S. Dept. of Commerce under federal award 60NANB20d001.
\bibliographystyle{IEEEtran}
\bibliography{IEEEabrv,references}

\end{document}

%% file: glossary.tex
\newacronym{scs}{SCS}{sub-carrier spacing}
\newacronym{ofdm}{OFDM}{Orthogonal Frequency Division Multiplexing}
\newacronym{fdm}{FDM}{Frequency Division Multiplexing}
\newacronym{cpu}{CPU}{Central Processing Unit}
\newacronym{tcp}{TCP}{Transmission Control Protocol}
\newacronym{tcpw}{TCPW}{TCP Wave}
\newacronym{lte}{LTE}{Long Term Evolution}
\newacronym{nr}{NR}{New Radio}
\newacronym{cwnd}{cWnd}{Congestion Window}
\newacronym{caa}{CAA}{Congestion Avoidance Algorithm}
\newacronym{bbr}{BBR}{Bottleneck Bandwidth and Round-trip propagation time}
\newacronym{nv}{NV}{New Vegas}
\newacronym{rtt}{RTT}{Round-Trip Time}
\newacronym{ietf}{IETF}{Internet Engineering Task Force}
\newacronym{rfc}{RFC}{Request For Comments}
\newacronym{gcc}{GCC}{GNU Compiler Collection}
\newacronym{tso}{TSO}{TCP Segmentation Offloading}
\newacronym{tsq}{TSQ}{TCP Small Queue}
\newacronym{gbr}{GBR}{Guaranteed Bit Rate}
\newacronym{nongbr}{non-GBR}{non-Guaranteed Bit Rate}
\newacronym{enb}{eNB}{Evolved Node B}
\newacronym{dpi}{DPI}{Deep Packet Inspection}
\newacronym{rlc}{RLC}{Radio Link Control}
\newacronym{bsr}{BSR}{Buffer Status Report}
\newacronym{qos}{QoS}{Quality of Service}
\newacronym{aqm}{AQM}{Active Queue Management}
\newacronym{rds}{RDS}{Radio Data Scheduler}
\newacronym{tc}{TC}{Traffic Control}
\newacronym{drb}{DRB}{Data Radio Bearer}
\newacronym{rnti}{RNTI}{Radio Network Temporary Identifier}
\newacronym{bql}{BQL}{Byte Queue Limits}
\newacronym{ue}{UE}{User Equipment}
\newacronym{am}{AM}{Acknowledged Mode}
\newacronym{epc}{EPC}{Evolved Packet Core}
\newacronym{cn}{CN}{Core Network}
\newacronym{gnb}{gNB}{next-Generation Node B}
\newacronym{ran}{RAN}{Radio Access Network}
\newacronym{3gpp}{3GPP}{3rd Generation Partnership Project}
\newacronym{5g}{5G}{fifth Generation}
\newacronym{dl}{DL}{DownLink}
\newacronym{ul}{UL}{UpLink}
\newacronym{tti}{TTI}{Transmission Time Interval}
\newacronym{sr}{SR}{Scheduling Request}
\newacronym{bsr}{BSR}{Buffer Status Report}
\newacronym{e2e}{E2E}{End-To-End}
\newacronym{embb}{eMBB}{enhanced Mobile BroadBand}
\newacronym{urllc}{URLLC}{Ultra-Reliable and Low-Latency Communications}
\newacronym{mmtc}{mMTC}{massive Machine Type Communications}
\newacronym{ul}{UL}{UpLink}
\newacronym{dl}{DL}{DownLink}
\newacronym{phy}{PHY}{Physical}
\newacronym{mac}{MAC}{Medium Access Control}
\newacronym{prb}{PRB}{Physical Resource Block}
\newacronym{rb}{RB}{Resource Block}
\newacronym{pps}{pps}{Packets Per Second}
\newacronym{cp}{CP}{Cyclic Prefix}
\newacronym{tbs}{TBS}{Transport Block Size}
\newacronym{tb}{TB}{Transport Block}
\newacronym{cb}{CB}{Code Block}
\newacronym{gtp}{GTP}{GPRS Tunneling Protocol}
\newacronym{sap}{SAP}{Service Access Point}
\newacronym{tm}{TM}{Transparent Mode}
\newacronym{um}{UM}{Unacknowledged Mode}
\newacronym{am}{AM}{Acknowledged Mode}
\newacronym{sm}{SM}{Saturation Mode}
\newacronym{sinr}{SINR}{Signal-to-Interference-plus-Noise Ratio}
\newacronym{rrc}{RRC}{Radio Resource Control}
\newacronym{tdma}{TDMA}{Time-Division Multiple Access}
\newacronym{ofdma}{OFDMA}{Orthogonal Frequency-Division Multiple Access}
\newacronym{rbg}{RBG}{Resource Block Group}
\newacronym{rb}{RB}{Resource Block}
\newacronym{dci}{DCI}{Downlink Control Information}
\newacronym{uci}{UCI}{Uplink Control Information}
\newacronym{ipat}{IPAT}{Inter-Packet Arrival Time}
\newacronym{pdsch}{PDSCH}{Physical Downlink Shared Channel}
\newacronym{pusch}{PUSCH}{Physical Uplink Shared Channel}
\newacronym{pucch}{PUCCH}{Physical Uplink Control Channel}
\newacronym{pdcch}{PDCCH}{Physical Downlink Control Channel}
\newacronym{pssch}{PSSCH}{Physical Sidelink Shared Channel}
\newacronym{pscch}{PSCCH}{Physical Sidelink Control Channel}
\newacronym{psbch}{PSBCH}{Physical Sidelink Broadcast Channel}
\newacronym{psfch}{PSFCH}{Physical Sidelink Feedback Channel}
\newacronym{tdd}{TDD}{Time Division Duplex}
\newacronym{fdd}{FDD}{Frequency Division Duplex}
\newacronym{rach}{RACH}{Random Access Channel}
\newacronym{cbr}{CBR}{Constant Bit Rate}
\newacronym{los}{LoS}{Line-of-Sight}
\newacronym{mcs}{MCS}{Modulation Coding Scheme}
\newacronym{bwp}{BWP}{Bandwidth Part}
\newacronym{cqi}{CQI}{Channel Quality Indicator}
\newacronym{bler}{BLER}{Block Error Rate}
\newacronym{tbler}{TBLER}{Transport Block Error Rate}
\newacronym{mi}{MI}{Mutual Information}
\newacronym{l2sm}{L2SM}{Link to System Mapping}
\newacronym{sliv}{SLIV}{Start and Length Indicator Value}
\newacronym{mmwave}{mmWave}{millimeter-wave}
\newacronym{pdu}{PDU}{Packet Data Unit}
\newacronym{ca}{CA}{Carrier Aggregation}
\newacronym{snr}{SNR}{Signal-to-Noise Ratio}
\newacronym{sinr}{SINR}{Signal to Interference-plus-Noise Ratio}
\newacronym{pdcp}{PDCP}{Packet Data Convergence Protocol}
\newacronym{sdap}{SDAP}{Service Data Adaptation Protocol}
\newacronym{sdu}{SDU}{Service Data Unit}
\newacronym{nas}{NAS}{Non-Access Stratum}
\newacronym{sme}{SME}{Small and Medium Enterprise}
\newacronym{rat}{RAT}{Radio Access Technology}
\newacronym{pgw}{PGW}{Packet data network GateWay}
\newacronym{sgw}{SGW}{Service GateWay}
\newacronym{ldpc}{LDPC}{low Density Parity Check}
\newacronym{cca}{CCA}{Clear Channel Assessment}
\newacronym{csmaca}{CSMA/CA}{Carrier Sense Multiple Access with Collision Avoidance}
\newacronym{ccm}{CCM}{Component Carrier Manager}
\newacronym{cam}{CAM}{Channel Access Manager}
\newacronym{cws}{CWS}{Contention Window Size}
\newacronym{ed}{ED}{Energy detection}
\newacronym{harq}{HARQ}{Hybrid Automatic Repeat Request}
\newacronym{ap}{AP}{Access Point}
\newacronym{laa}{LAA}{Licensed-Assisted Access}
\newacronym{lbt}{LBT}{Listen-Before-Talk}
\newacronym{mcot}{MCOT}{Maximum Channel Occupancy Time}
\newacronym{nru}{NR-U}{New Radio-based access to Unlicensed spectrum}
\newacronym{mcot}{MCOT}{Maximum Channel Occupancy Time}
\newacronym{cot}{COT}{Channel Occupancy Time}
\newacronym{sta}{STA}{Station}
\newacronym{wlans}{WLANs}{Wireless Local Area Networks}
\newacronym{ss/pbch}{SS/PBCH}{Synchronization Signal/Physical Broadcast Channel}
\newacronym{prach}{PRACH}{Physical Random Access Channel}
\newacronym{ocb}{OCB}{Occupied Channel Bandwidth}
\newacronym{sps}{SPS}{Semi-Persistent Scheduling}
\newacronym{eps}{EPS}{Evolved Packet System}
\newacronym{lcid}{LCID}{Logical Channel Identifier}
\newacronym{lc}{LC}{Logical Channel}
\newacronym{csi}{CSI}{Channel State Information}
\newacronym{csirs}{CSI-RS}{Channel State Information Reference Signal}
\newacronym{ptrs}{PT-RS}{Phase Tracking Reference Signal}
\newacronym{prsvp}{$P_\text{rsvp}$}{Resource Reservation Period}
\newacronym{slrrc}{SLRRC}{Sidelink Resource Reselection Counter}
\newacronym{rv}{RV}{Redundancy Version}
\newacronym{ndi}{NDI}{New Data Indicator}
\newacronym{rsrp}{RSRP}{Reference Signal Received Power}
\newacronym{re}{RE}{Resource Element}
\newacronym{dmrs}{DMRS}{Demodulation Reference Signals}
\newacronym{prose}{ProSe}{Proximity Services}
\newacronym{si}{SI}{Study Item}
\newacronym{wi}{WI}{Work Item}
\newacronym{tr}{TR}{Technical Report}
\newacronym{ts}{TS}{Technical Specification}
\newacronym{d2d}{D2D}{Device-to-Device}
\newacronym{eutra}{E-UTRA}{Evolved Universal Terrestrial Radio Access}
\newacronym{v2x}{V2X}{Vehicular-to-everything}
\newacronym{cv2x}{C-V2X}{Cellular V2X}
\newacronym{v2v}{V2V}{Vehicle-to-Vehicle}
\newacronym{v2p}{V2P}{Vehicle-to-Pedestrian}
\newacronym{v2i}{V2I}{Vehicle-to-Infrastructure}
\newacronym{v2n}{V2N}{Vehicle-to-Network}
\newacronym{5gaa}{5GAA}{5G Automotive Association}
\newacronym{dsrc}{DSRC}{Dedicated Short Range Communications}
\newacronym{3gpp}{3GPP}{Third Generation Partnership Project}
\newacronym{its}{ITS}{Intelligent Transport System}
\newacronym{ie}{IE}{Information Element}
\newacronym{kpi}{KPI}{Key Performance Indicator}
\newacronym{pdb}{PDB}{Packet Delay Budget}
\newacronym{pir}{PIR}{Packet Inter-reception Delay}
\newacronym{sci}{SCI}{Sidelink Control Information}
\newacronym{tft}{TFT}{Traffic Flow Template}
\newacronym{um}{UM}{Unacknowledged Mode}
\newacronym{cdf}{CDF}{Cumulative Density Function}
\newacronym{mimo}{MIMO}{Multiple-Input Multiple-Output}

%% file: abstract.tex
In this paper, we use a New Radio (NR) Vehicular-to-everything (V2X) standard compliant simulator based on \textit{ns-3}, to study the impact of NR numerologies on the end-to-end performance. In particular, we focus on NR V2X Mode 2, used for autonomous resource selection in out-of-coverage communications, and consider the two key procedures defined in 3GPP: sensing and non-sensing based resource selection. We pay particular attention to the interplay between the operational numerology and the resource selection window length, a key parameter of NR V2X Mode 2. The results in a standard-compliant, end-to-end simulation platform show that in all cases, for basic service messages, a higher numerology is beneficial because of different reasons, depending on the way the resource selection window length is established.

%% file: introduction.tex
Building upon what has been standardized for \gls{d2d} and \gls{lte} \gls{cv2x}, the \gls{3gpp} has continued the standardization efforts on \gls{v2x} communications in Release 16 and 17~\cite{RP-190766}, for \gls{nr} access. The idea is to enable a wide range of \gls{v2x} applications with different quality of service requirements and support scenarios with high vehicular densities~\cite{TR22886, 8080373}. A support for diverse applications and use cases is possible in \gls{nr} \gls{v2x} because of the flexible framework inherited by the \gls{nr} technology and the recent progresses envisioned in \gls{nr} \gls{v2x}. In particular, NR provides wide bandwidth support in various frequency ranges, flexible frame structure with reduced transmission time intervals (TTIs) (using multiple numerologies), support for massive \gls{mimo} systems, high modulation orders, and advanced channel coding~\cite{parkvall:17}. All these new features and functionalities intrinsically contribute to increase the data rate, reduce the latency, and improve the spectral efficiency of \gls{v2x} communication systems. In addition, new enhancements and key procedures have been defined for \gls{nr} \gls{v2x}, specifically designed to improve the reliability of \gls{v2x} communications systems. For example, new communication types (unicast and groupcast), a new feedback channel, the support of feedback-based retransmissions, and new resource allocation and scheduling mechanisms~\cite{TR38885}.

NR V2X defines two resource allocation modes for sidelink communications, one centralized (Mode 1) and one distributed (Mode 2)~\cite{TR38885}. 
These two NR V2X modes are similar to LTE C-V2X Modes 3 and 4, respectively. NR V2X Mode 1 is a centralized scheduling approach, in which the resource allocation is managed by the base station (gNB in NR) and applies to scenarios in which the various users (UEs) are inside the coverage of the gNB (i.e., in-coverage scenarios). On the other hand, NR V2X Mode 2 is a distributed scheduling approach in which the resource allocation is carried out by the UEs themselves, with no need to be in the coverage area of the gNB (i.e., it supports out-of-coverage communications). In this paper, we focus on NR V2X Mode 2 with periodic traffic.

Resource reservation for NR V2X Mode 2 under periodic traffic mostly reuses the \gls{lte} \gls{cv2x} sidelink Mode 4 long-term sensing-based algorithm, which exploits the periodicity and fixed-size assumption of basic safety messages. In addition to the long-term sensing-based resource selection, NR V2X Mode 2 also supports a non-sensing resource selection to reduce the complexity of the UE and the power consumption~\cite{RP-193231}. The difference between sensing and non-sensing based resource selections is that, before selecting the resources from the total available ones, the sensing-based procedure filters those slots which are in use by other UEs, using sensing information. On the other hand, the non-sensing-based procedure does not use the sensing information and directly selects the resources from the total available ones. 

While \gls{lte} C-\gls{v2x} has been widely studied analytically and through simulations by academia and industry~\cite{8581518,8080373,9133075}, the studies on \gls{nr} \gls{v2x} have just started. Overviews of the standardization activities and NR V2X design principles in \gls{3gpp} Release 16 are provided in~\cite{9214394,9063405,8998153,9345798}. 
However, few of these works discuss simulation studies and to the best of the authors' knowledge none of them is based on Release 16 \gls{nr}-compliant \gls{v2x} simulation models, because the standardization has recently been completed. In addition, a key challenge to evaluate performance of \gls{nr} \gls{v2x} is that, despite the set of simulation results by industry and in literature, the simulators are not publicly available.

In this paper, we consider an extension of the open source, end-to-end, ns-3 5G-LENA simulator~\cite{PATRICIELLO2019101933}, which we have developed to support \gls{nr} \gls{v2x} capabilities~\cite{zali}. Building upon such a standard-compliant simulation platform, we focus on assessing the impact of the NR numerologies on the NR V2X Mode 2, and to understand how it affects the PHY and MAC layers, from an end-to-end perspective. In~\cite{numero:1}, the impact of the NR numerology on the \gls{v2x} autonomous sidelink mode is assessed, but simulations are carried out over an \gls{lte} C-\gls{v2x} simulator. In~\cite{numero:2}, authors study the use of \gls{nr} numerology in a \gls{v2v} scenario characterized by different vehicle speeds. However, the simulations are conducted on an \gls{lte} PHY layer simulator. Both,\cite{numero:1} and \cite{numero:2}, concluded that a higher numerology reduces the TTI length and so it is beneficial to improve the reception ratio and delay performance. In this paper, we want to verify such conclusions from an end-to-end perspective by considering NR V2X-compliant PHY and MAC layers and both sensing and no-sensing based resource selection procedures. 

This paper is structured as follows. Sec.~\ref{sec:2} overviews NR V2X Mode 2 defined in \gls{3gpp}, focusing on the details of the resource selection procedure. In Sec.~\ref{sec:3}, we elaborate on the impact of the NR numerologies on the PHY layer and the MAC resource selection procedure of NR V2X Mode 2. Sec.~\ref{sec:4} presents the simulation scenario and the simulation results for different numerologies and resource selection window lengths. Finally, Sec.~\ref{sec:conclusions} concludes the paper.

%% file: sec-2.tex
NR V2X Mode 2 considers sensing-based Semi-Persistent Scheduling (SPS) for periodic traffic. This is defined as a distributed scheduling protocol to autonomously select radio resources, in a similar way to what is already considered for LTE C-V2X Mode 4. The sensing procedure takes advantage of the periodic and predictable nature of V2X basic service messages. In particular, sensing-based SPS UEs reserve subchannels in the frequency domain for a random number of consecutive periodic transmissions in time domain. The number of slots for transmission and retransmissions within each periodic resource reservation period depends on the number of blind retransmissions (if any) and the resource selection procedure. The number of reserved subchannels per slot depends on the size of data to be transmitted. 

\begin{figure*}[!t]
  \centering
  \includegraphics[width=0.8\linewidth]{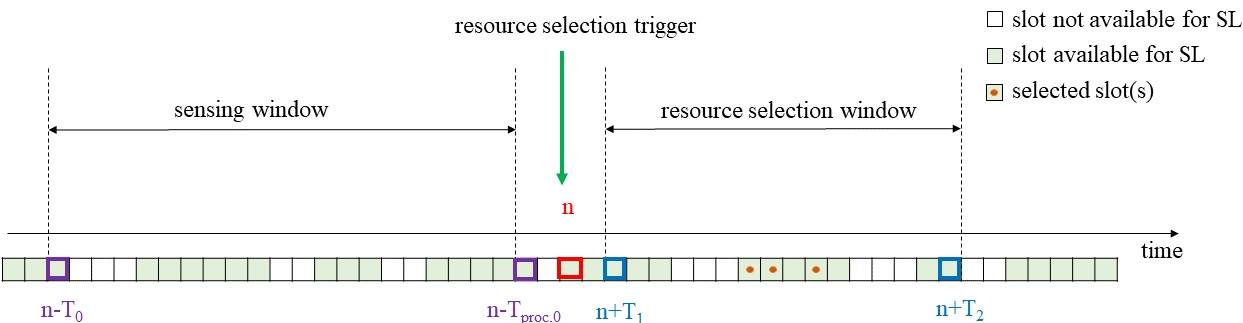}
  \caption{NR V2X Mode 2 resource selection procedure. $T_\text{0}=20$ slots, $T_\text{proc,0}=2$ slots, $T_\text{1}=2$ slots, and $T_\text{2}=17$ slots.} 
  \label{fig:sensing}
  \vspace{-0.3cm}
\end{figure*}

\subsubsection{Sensing-based resource selection procedure}
\label{res-selec}
The sensing-based resource selection procedure is composed of two stages: 1) a sensing procedure and 2) a resource selection procedure~\cite{TS38321}. 
The sensing procedure is in charge of identifying the resources that are candidate during the resource selection. In particular, it is based on the decoding of the 1st-stage-Sidelink Control Information (1st-stage-SCI) received from the surrounding UEs and on sidelink power measurements~\cite{TS38214}. The sensing procedure is performed during the so-called \textit{sensing window}, defined by the pre-configured parameter $T_\text{0}$ and a UE-specific parameter $T_\text{proc,0}$ that accounts for the time required to complete SCIs decoding and possibly perform measurements for the sensing procedure. Specifically, if at time $n$ the sensing-based resource selection is triggered, the UE will consider the sidelink measurements performed during the interval [$n-T_\text{0}, n-T_\text{proc,0}$).

\begin{figure*}[!t]
  \centering
  \includegraphics[width=0.8\linewidth]{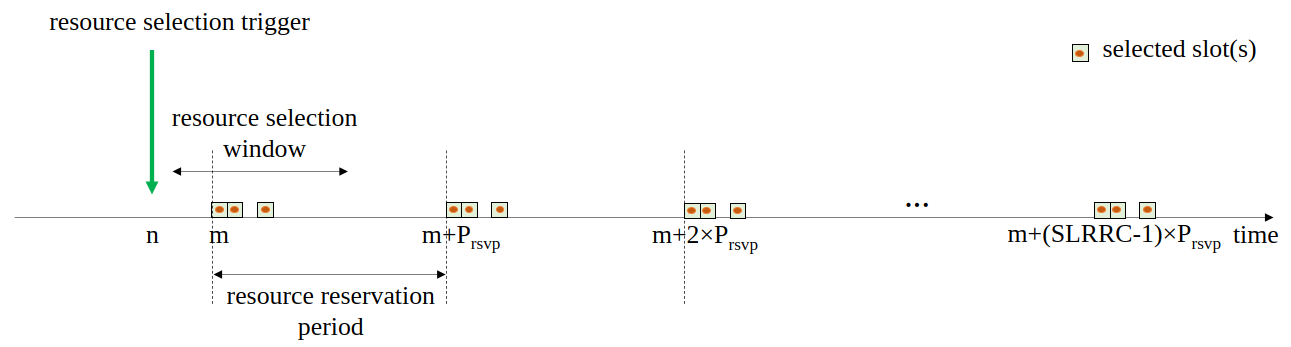}
  \caption{NR V2X Mode 2 semi-persistent scheduling.} 
  \label{fig:sps}
  \vspace{-0.3cm}
\end{figure*}

Based on the information extracted from the sensing, the resource selection procedure determines the resource(s) for sidelink transmissions~\cite{TS38321}. For that, another window called the \textit{resource selection window}, is defined. The resource selection window is bounded by the interval [$n+T_\text{1}, n+T_\text{2}$], where $T_\text{1}$ and $T_\text{2}$ are two parameters that are determined by the UE implementation~\cite{TS38214}. $T_\text{2}$ depends on the \gls{pdb} and on an RRC pre-configured parameter called $T_\text{2,min}$. In case, $\text{PDB}>T_\text{2,min}$, $T_\text{2}$ is determined by the UE implementation and must meet the following condition: $T_\text{2,min}<=T_\text{2}<=\text{PDB}$. In case $\text{PDB}\le T_\text{2,min}$, $T_\text{2}=\text{PDB}$. $T_\text{1}$ is selected so that $T_\text{proc,1}<=T_\text{1}$, where $T_\text{proc,1}$ is the time required to identify the candidate resources and select a subset of resources for sidelink transmission. The resource selection procedure is composed of two steps. First, the candidate resources within the resource selection window are identified. 
A resource is indicated as non-candidate if an SCI is received on that slot or the corresponding slot is reserved by a previous SCI, and the associated sidelink measurement is above a threshold~\cite{TS38214}. To proceed with the second step, the resulting set of candidate resources within the resource selection window should be at least a $X$~\% of the total resources within the resource selection window. The value of $X$ is configured by RRC and can be $20$~\%, $35$~\% or $50$~\%.  If this condition is not met, the threshold is increased by 3 dB and the procedure is repeated. Second, the transmitting UE performs the resource selection from the identified candidate resources (which may include initial transmissions and retransmissions). For that, a randomized resource selection from the identified candidate resources in the resource selection window is supported.

To exclude resources from the candidate pool based on sidelink measurements in previous slots, the resource reservation period is introduced, which is communicated by the neighbour UEs through the 1st-stage-SCI. 

The UE that performs the resource selection uses this periodicity (if included in the decoded SCI) and assumes that the neighbor UE will do periodic transmissions with such a periodicity during $Q$ periods. This allows to identify and exclude the non-candidate resources in the resource selection window. According to~\cite{TS38214}, $Q=\lceil{\frac{T_\text{scal}}{P_\text{rsvp}}}\rceil$, where $P_\text{rsvp}$ refers to the resource reservation period used by the neighbouring UEs, and $T_\text{scal}$ corresponds to $T_\text{2}$ converted to units of ms~\cite{TS38214}.

As previously mentioned, NR V2X also supports a non-sensing based resource selection~\cite{TS38321}. In this case, the sensing procedure is omitted, and all the resources within the resource selection window that are part of the resource pool for sidelink are candidates for random selection.

Fig.~\ref{fig:sensing} shows the resource selection procedure in NR V2X Mode 2. The figure illustrates the sensing window and resource selection window, with an example that uses $T_\text{0}=20$ slots, $T_\text{proc,0}=2$ slots, $T_\text{1}=2$ slots, and $T_\text{2}=17$ slots. Note that, with this configuration, the resulting resource selection window length is $T_\text{2}-T_\text{1}+1=16$ slots. Once the resource selection is triggered at time $n$, based on the measurements in the sensing window, the MAC scheduler determines the transmission resources within the resource selection window, which can be used for different MAC PDUs or to perform blind retransmissions.

\subsubsection{Semi-persistent scheduling}
\label{sps}
Once one or multiple resources are selected, the UE will consider periodic transmissions using SPS. The transmission interval is defined by the \gls{prsvp}, which is pre-configured by RRC and can take predefined values between 1 ms and 1000 ms~\cite{TS38321}. \gls{prsvp} value is included in the 1st-stage-SCI, to allow other UEs to estimate which resources are reserved in the future based on SCI decoding. After using the resource for the number of transmissions equal to the \gls{slrrc}, a resource reselection is triggered. Whether to reselect or not, depends on the configured probability of keeping the current resources, known as ``probability of resource keep''. In particular, once \gls{slrrc} reaches zero, the UE either keeps the previous selection or selects new resources based on the pre-configured probability value. The value of \gls{slrrc} is randomly selected from the interval [$5, 15$] for \gls{prsvp} $\ge 100$~ms. For \gls{prsvp} $< 100$ ms, the value of \gls{slrrc} is randomly selected from the interval $\lceil{5\times\frac{100}{\max(20,P_\text{rsvp})}, 15\times\frac{100}{\max(20,P_\text{rsvp})}\rceil}$~\cite{TS38321}. The standard also defines the maximum number of times that the same resource can be used for SPS through $C_\text{resel}=10\times\text{SLRRC}$, after which the resource reselection has to be triggered, independently of the probability of resource keep.
An illustration of the SPS procedure for NR V2X Mode 2 is shown in Fig.~\ref{fig:sps}. In the example, three resources are selected within the resource selection window ($m$ in the figure is the slot index of the first selected resource), and these allocations are repeated every \gls{prsvp} for \gls{slrrc} times. Once the three transmissions in the interval starting at $m+(\text{SLRRC}-1)\times P_\text{rsvp}$ have been carried out, either the same selection is kept or a new resource selection procedure is triggered, based on the probability of resource keep.

%% file: sec-3.tex
With flexibility in mind, \gls{nr} includes multiple numerologies, each being defined by a \gls{scs} and a \gls{cp}.
The supported numerologies ($\mu$) in NR V2X can take values from 0 to 3 and specify an \gls{scs} of $15{\times}2^\mu$ kHz and a slot length of $1/2^\mu$ ms~\cite{TS38211}. In particular, $\mu=0,1,2$ are supported in frequency range 1 (FR1, in sub 6 GHz bands) and $\mu=2,3$ are supported in frequency range 2 (FR2, in millimeter-wave bands). The standardization on NR V2X has first focused (within Release 16) in FR1, and so, supporting $\mu=0,1,2$.

\begin{figure*}[!t]
  \centering
  \includegraphics[width=0.8\linewidth]{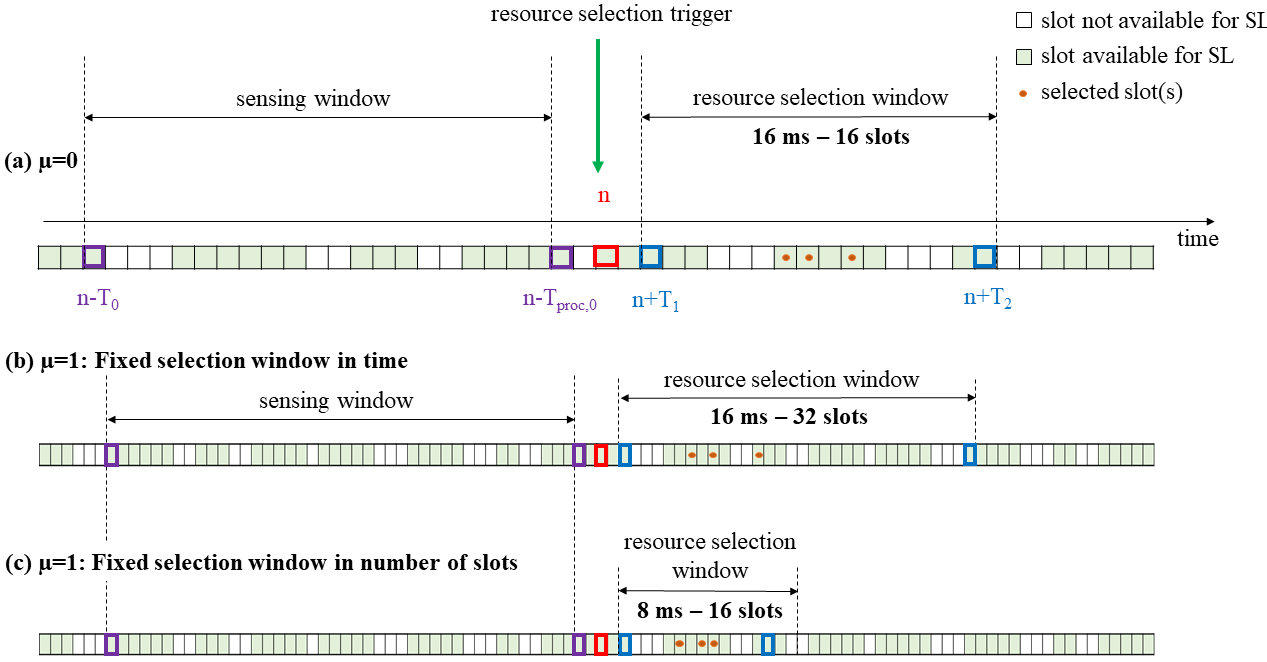}
  \caption{Impact of numerologies on NR V2X Mode 2 resource selection procedure. Example with $T_\text{0}=20$ slots, $T_\text{proc,0}=2$ slots, $T_\text{1}=2$ slots. (a) $\mu$=0 and $T_\text{2}=17$ slots, (b) $\mu$=1 and $T_\text{2}=33$ slots, (c) $\mu$=1 and $T_\text{2}=17$ slots.} 
  \label{fig:sensingMu}
  \vspace{-0.2cm}
\end{figure*}

The operational numerology affects the NR V2X frame structure, including the slot length (and so the TTI length) and the \gls{rb} width, as well as the processing delays~\cite{8514979}. Furthermore, in the case of NR V2X, it also affects the sizes of the sensing and resource selection windows, previously introduced in Sec.~\ref{sec:2}, and so the resource selection procedure at the \gls{mac} layer itself. In NR V2X standard, $T_\text{0}$ is defined in ms, and so its actual length in time (in ms) will be the same over different numerologies, but higher numerologies would consider more slots within the sensing window. On the other hand, $T_\text{2}$ (which defines the end of the selection window) is defined in a number of slots. Even though, as previously mentioned, the $T_\text{2}$ determination is up to the UE implementation and depends on the \gls{pdb} that is defined in ms. If the \gls{pdb} is tight (i.e., $\text{PDB}\le T_\text{2,min}$) then $T_\text{2}$ will be set in slots, such that, it satisfies the required \gls{pdb} irrespective of the numerology used. Therefore, the resulting resource selection window's length will have the same duration in time (in ms) for different numerologies. However, under a more relaxed \gls{pdb} (i.e., $\text{PDB} > T_\text{2,min}$), $T_\text{2}$ can be set to any number slots specified by the \gls{3gpp} in TS38.331. In this case, different numerologies lead to resource selection windows of the same number of slots but different lengths (in ms).

To illustrate this, in Fig.~\ref{fig:sensingMu} we show the NR V2X Mode 2 resource selection procedure in three different cases: (a) $\mu=0$ and $T_\text{2}=17$, (b) $\mu=1$ and $T_\text{2}=33$, (c) $\mu=1$ and $T_\text{2}=17$. In these three figures, we use $T_\text{0}=20$ slots, $T_\text{proc,0}=2$ slots, $T_\text{1}=2$ slots. In (a), the resource selection window length results in 16 ms and 16 slots. In (b), the resource selection window length results in 16 ms and 32 slots, so that it has the same length in ms as for case (a). Finally, in (c), the resource selection window length results in 8 ms and 16 slots, so that it has the same length in number of slots as for case (a). 
Let us note that the same behaviour can be extended for the case of $\mu=2$, for which the slot is halved as compared to $\mu=1$.

As it can be observed, when maintaining the resource selection window length in time (in ms) over different numerologies, we have more slots within a resource selection window to choose from if a larger numerology is used (see Fig.~\ref{fig:sensingMu}.(a) and Fig.~\ref{fig:sensingMu}.(b)). On the other hand, when maintaining the resource selection window length in the number of slots over different numerologies, then we get the same number of slots within a resource selection window, but the actual length of the window (in ms) is reduced if a larger numerology is used (see Fig.~\ref{fig:sensingMu}.(a) and Fig.~\ref{fig:sensingMu}.(c)). Therefore, different behaviours can be expected when comparing the numerology's impact in the NR V2X system, depending on how the resource selection window length is set for the comparison (i.e., fixed in time or in the number of slots). 

%% file: sec-4.tex
\begin{figure}[!b]
  \vspace{-0.2cm}
  \includegraphics[width=1\linewidth]{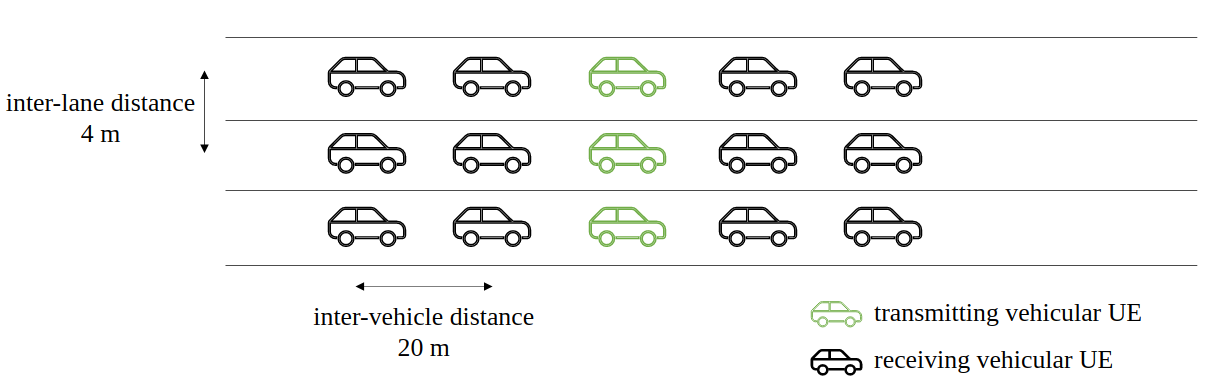}
  \caption{Highway scenario with 3 lanes and 5 vehicular UEs per lane moving at a speed of 140 km/h.}
  \label{fig:scenario}
\end{figure}

For the end-to-end evaluation, we model a \gls{v2x} highway scenario, as described by \gls{3gpp}~\cite{TR37885}. The standard deployment consists of multiple highway lanes with an inter-lane distance of 4 m. A different number of vehicles with an inter-vehicle distance of 20 m can be dropped within each lane. In our simulation, we consider 3 lanes in which 5 vehicular UEs that move in the same direction are dropped per lane by following Option A in~\cite{TR37885}. No cluster dropping is used, and all the UEs are passenger vehicles with an antenna height of 1.6 m. Without loss of generality, we focus on the platooning use case and assume that the center vehicle in each lane is a transmitter, while the other vehicles are receivers, as shown in Fig.~\ref{fig:scenario}. Each transmitter periodically transmits a 200 bytes long packet every 100 ms (i.e., 16 kb/s data rate) over a 5.9 GHz band with 40 MHz bandwidth~\cite{TR38885}. It is worth mentioning that, here, we focus on an out-coverage-scenario, and therefore gNBs are not simulated. Table~\ref{sim-params-table} summarizes all the simulation parameters.

\input{simulation-param.tex}

We organize the simulation campaign in two categories: 1) we fix the resource selection window in time (ms) and 2) we fix the resource selection window in number of slots. For each category, we study the impact of using different NR V2X numerologies ($\mu$) over sensing and no-sensing resource selection procedures. In particular, we consider the three numerologies supported by the standard in FR1: $\mu=0$ (15 kHz \gls{scs}), $\mu=1$ (30 kHz \gls{scs}), and $\mu=2$ (60 kHz \gls{scs}). They are displayed in the figures' legends as mu-$0$, mu-$1$, and mu-$2$, respectively. To obtain statistically significant results, 50 random channel realizations are performed for each simulation. The single simulation has a duration of 10 s, where the application of each transmitter starts at a random time within an interval of 100 ms. Finally, for the performance evaluation we use two \glspl{kpi}:
\begin{itemize}
    \item \gls{pir}: average interval of time elapsed between two successful packet receptions, measured at the application layer, for each transmit-receive pair~\cite{TR37885}.
    \item The percentage of simultaneous \gls{pssch} transmissions over the total number of \gls{pssch} transmissions by all the transmitting UEs (trace triggered from the MAC layer).
\end{itemize}

In the following subsections, for each category, we present the \gls{cdf} of the above \glspl{kpi} and discuss the obtained results based on them.

\begin{figure*}[!t]
  \centering
  \subfloat(a){\includegraphics[width=0.47\linewidth]{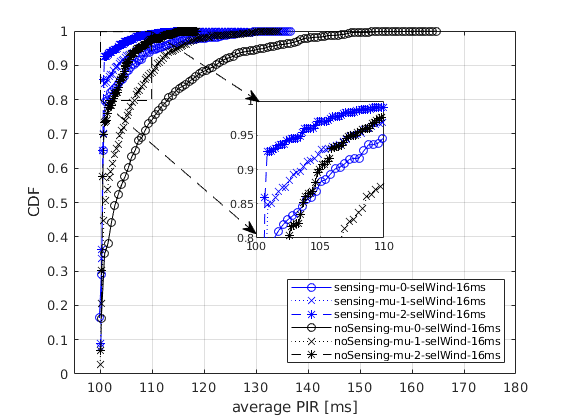}}
  \subfloat(b){\includegraphics[width=0.46\linewidth]{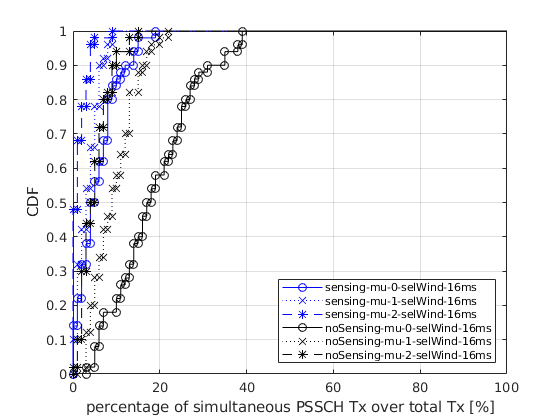}}\\
  \caption{Impact of NR V2X numerology ($\mu$) when the resource selection window length is fixed in time. (a) PIR (ms), (b) percentage of simultaneous PSSCH transmissions over total transmissions (\%).}
  \vspace{-0.2cm}
  \label{fig:t2-fix-in-time}
\end{figure*}

\begin{figure*}[!t]
  \centering
  \subfloat(a){\includegraphics[width=0.47\linewidth]{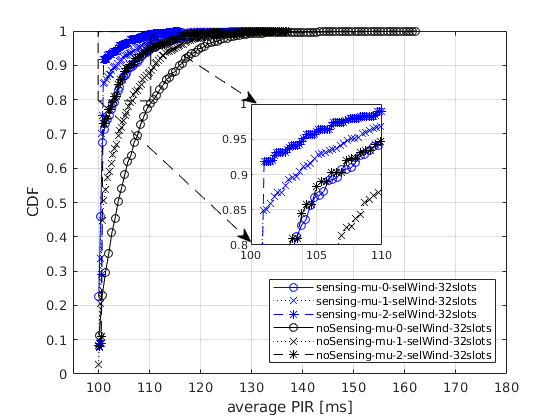}}
  \subfloat(b){\includegraphics[width=0.47\linewidth]{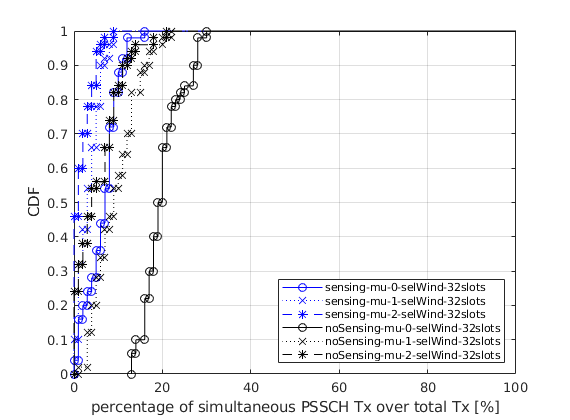}}\\
  \caption{Impact of NR V2X numerology ($\mu$) when the resource selection window length is fixed in number of slots. (a) PIR (ms), (b) percentage of simultaneous PSSCH transmissions over total transmissions (\%).}
  \vspace{-0.2cm}
  \label{fig:t2-fix-in-slots}
\end{figure*}

\subsection{Fixed resource selection window in time}
\label{sec:fixed-in-time}
As explained in Sec.~\ref{sec:2}, according to the \gls{3gpp} standard, the size of the resource selection window should be configured by taking into account the \gls{pdb}. Following this requirement, in the first simulation campaign, we configure the size of the resource selection window in the number of slots, so that it is 16 ms long in time for each considered numerology, which fulfills a \gls{pdb} requirement of 20 ms. Specifically, the parameter $T_\text{2}$ is set to 17, 33, and 65 slots for $\mu=0$, $\mu=1$, and $\mu=2$, respectively. Fig.~\ref{fig:t2-fix-in-time} shows the CDF statistics of (a) the PIR and (b) the percentage of simultaneous \gls{pssch} transmissions.

In Fig.~\ref{fig:t2-fix-in-time}.(b), as expected, the sensing based resource selection reduces the number of simultaneous \gls{pssch} transmissions compared to noSensing. That is, for a fixed $\mu$, the sensing based resource selection effectively decreases the probability of collision, and consequently, an improvement in PIR is observed in Fig.~\ref{fig:t2-fix-in-time}.(a). For example, using $\mu=0$, in the considered scenario, the median for the simultaneous \gls{pssch} transmissions is at 5~\% with the sensing-based resource selection compared to 18~\% with the noSensing-based procedure (see Fig.~\ref{fig:t2-fix-in-time}.(b)). Similarly, the improvement in terms of PIR is that for the 58.4~\% of the cases, we get the ideal PIR of 100 ms with sensing, while this result is reduced to around an 6.8~\% in case of noSensing. Thus, these results validate the belief about the sensing and the gains it brings over the noSensing-based resource selection.

Furthermore, for our chosen simulation parameters, we also observe that irrespective of the resource selection procedure, i.e., sensing or noSensing, using higher numerology always improves the performance in terms of PIR and number of simultaneous \gls{pssch} transmissions. The reason is that in \gls{nr} systems the slot length in time is inversely proportional to the \gls{scs} (i.e., higher \gls{scs} implies lower slot duration) in frequency. In this simulation campaign, the resource selection window size in time is the same for all the tested numerologies, i.e., 16 ms. However, by increasing the numerology, we basically double the size of the window in slots. For example, the resource selection window should consist of 16 slots for it to be 16 ms in time when using $\mu=0$. On the other hand, for the same duration of the resource selection window, the number of slots are 32 and 64 for $\mu=1$ and $\mu=2$, respectively. It means, with higher numerology, a transmitting UE would have more slots to choose from at the time of resource selection. This results in a lower number of simultaneous \gls{pssch} transmissions and the lower PIR for $\mu=1$ and $\mu=2$ compared to $\mu=0$, as shown in Fig.~\ref{fig:t2-fix-in-time}. In fact, in Fig.~\ref{fig:t2-fix-in-time}, for the same reason, we observe that with $\mu=2$ and when using noSensing, we are able to achieve similar performance to the one obtained using sensing with $\mu=1$. In summary, the availability of a higher number of slots to choose from reduces the probability of selecting the same resources by the transmitting vehicular UEs; hence, we see a better performance in terms of both the KPIs.

\subsection{Fixed resource selection window in number of slots}
\label{sec:fixed-in-slots}
The conclusion made in the previous subsection, i.e., higher number of slots are beneficial to reduce collisions, leads us to an interesting question: Would we achieve a better performance using higher numerology, if, the number of slots in the resource selection window is the same for all the tested numerologies? In this simulation campaign, we configure a resource selection window of 32 slots for all the tested numerologies. To do so, we set the parameter $T_\text{2}$ to 33 slots.

As shown in Fig.~\ref{fig:t2-fix-in-slots}, similar to our first simulation campaign, here the sensing based resource selection outperforms the noSensing one for both of the KPIs considered in this study. In particular, with sensing in 50~\% of the cases, we get the PIR of 100 ms compared to 23.3~\% with noSensing, as shown in Fig.~\ref{fig:t2-fix-in-slots}.(a). For the number of simultaneous \gls{pssch} transmissions, for the sensing, there is only 7~\% overlapping transmission in 50~\% of the cases. On the other hand, it increases to 20~\% for the noSensing resource selection.

Regarding the effect of the tested numerologies, once again, we observe that the use of higher numerology in our scenario improves the performance for the two considered metrics: it decreases the number of simultaneous \gls{pssch} transmissions, which leads to reduced PIR among the packets at the application layer. This outcome initially seems a bit counter-intuitive because in this campaign, for all the numerologies, the resource selection window consists of the same number of slots (i.e., 32). Therefore, the probability of selecting the same resource by the UEs is the same, irrespective of the numerology used. Thus, one could have expected to get a similar performance with the three tested numerologies when using sensing or noSening resource selection. However, this effect has not been observed. The reason, once more, lies in the concept of having a shorter slot length in time when increasing \gls{scs} in the \gls{nr} systems. With a shorter slot length, the duration of the resulting resource selection window (set to 32 slots) reduces in time. For example, a 32 slots resource selection window spans over 32 ms with $\mu=0$, 16 ms with $\mu=1$, and 8 ms with $\mu=2$. Interestingly, this reduction in the length of the resource selection window in ms decreases the probability of overlapping between UEs' selection window, because of the fixed resource reservation interval of 100 ms. For example, when using $\mu=0$ and 32 slots for a resource selection window, the windows for two UEs could overlap in time so that the end slots of the first resource selection window overlap with the starting slots of the second window. In such a case, increasing the numerology, e.g., to $\mu=1$, will decrease the windows' duration to half, i.e., 16 ms in time, thus reducing (or even fully avoiding) the overlapping. In other words, the first resource selection window will end 16 ms earlier, compared to $\mu=0$, thus, avoiding overlapping with the start of the second window. Consequently, it reduces the probability of collisions between the resource selection windows of these UEs.

%% file: simulation-param.tex
\begin{table}[t]
\caption{Simulation parameters.}
\label{sim-params-table}
\centering
\begin{tabular}{|p{0.2\textwidth}|p{0.2\textwidth}|}
\hline
\rowcolor[HTML]{C0C0C0} Parameter & Value \\
\hline
\textbf{Deployment and propagation parameters:} & \\
\hline
Channel model & 3GPP Highway\\
\hline
Deployment & 3 lanes, 5 vehicles per lane\\
\hline
Carrier frequency & 5.89 GHz\\
\hline
Channel bandwidth & 40 MHz\\
\hline
Noise power spectral density & -174 dBm/Hz\\
\hline
UE antenna height & 1.6 m\\
\hline
UE speed & 140 km/h\\
\hline
\textbf{Traffic parameters:} & \\
\hline
Application packet size & 200 Bytes\\
\hline
Inter-Packet interval & 100 ms\\
\hline
Application datarate & 160 kb/s\\
\hline
\textbf{Device parameters:} & \\
\hline
UE antennas & uniform planar array 4x2\\
\hline
UE transmit power & 23 dBm \\
\hline
UE noise figure & 5 dB\\
\hline
\textbf{NR V2X parameters:} & \\
\hline
Numerology ($\mu$) & 0, 1, 2\\
\hline
TDD pattern & $[\text{D D D S U U U U U U}]$\\
\hline
Sidelink bitmap	& $[\text{1 1 1 1 1 1 0 0 0 1 1 1}]$\\
\hline
Subchannel size ($N$) & 50 RBs \\
\hline
PSCCH symbols & 1 \\ 
\hline
PSSCH symbols & 12 \\ 
\hline
Link adaptation & fixed MCS \\ 
\hline
MCS index PSSCH & 14 from MCS Table2 \\
\hline
MCS index PSCCH & 0 from MCS Table2  \\
\hline
Error model & NR PHY abstraction based on EESM~\cite{lagen20} for PSSCH and PSCCH \\
\hline
Number of PSSCH transmissions ($N_\text{\gls{pssch},maxTx}$)   & 5 \\ 
\hline
HARQ combining method    & HARQ incremental redundancy \\ 
\hline
MAC resource selection   & sensing-based, no-sensing-based  \\
\hline
RLC mode & RLC-UM \\
\hline
RLC buffer size & 999999999 Bytes\\
\hline
\textbf{NR V2X Mode 2 parameters:}  & \\ 
\hline
Sensing window ($T_\text{0}$)	&	100 ms \\
\hline
Selection window ($T_\text{2}$) &	Fixed selection window in slots: \newline 33 slots (for all $\mu$) \newline Fixed selection window in time: \newline 17 slots for ($\mu = 0$) \newline 33 slots for ($\mu = 1$) \newline 65 slots for ($\mu = 2$)\\
\hline
$T_\text{1}$ & 2 slots \\
\hline
$T_\text{proc,0}$ & 2 slots\\
\hline
Percentage of resources must be selected in a selection window & 20~\% \\
\hline
Max num per reserve ($N_{\max\_\text{reserve}}$) & 3\\
\hline
Probability of resource keep & 0\\
\hline
Resource reservation period ($P_\text{rsvp}$) & 100 ms\\
\hline
Sensing RSRP threshold & -128 dBm\\
\hline
\end{tabular}
\vspace{-0.2cm}
\end{table}

%% file: conclusions.tex
In this paper, we analyzed the impact of the numerology and the size of the resource selection window on the \gls{pir} and the percentage of simultaneous transmissions in \gls{nr} \gls{v2x} Mode 2, employing sensing and non-sensing based resource selection. To do so, an extensive simulation analysis is carried out by extending the open-source ns-3 5G-LENA simulator following \gls{3gpp} standard on \gls{nr} sidelink communications. Based on the analysis using the simulation scenario, we conclude the following:
\begin{itemize}
    \item For a particular numerology, the sensing based resource selection, irrespective of the resource selection window length, outperforms a non-sensing based resource selection procedure.
    \item When the resource selection window size, expressed in time (i.e., ms) or in the number of slots, is the same for all the numerologies, a higher numerology results in a better performance for both sensing and non-sensing based resource selection.
    \item Using a higher numerology with non-sensing based resource selection provides an acceptable and close enough performance to the one obtained using sensing. 
\end{itemize}
To summarize, beside verifying the findings in the literature on the use of higher numerology with sensing based resource selection, this paper also provides insights for the non-sensing based resource selection procedure currently being studied for future \gls{nr} \gls{v2x} standard. Moreover, the results show the possibility to use a non-sensing based resource selection with an acceptable reduction in the performance, which presents a very interesting trade-off in terms of reducing the energy consumption and the complexity of a device versus the gain obtained using sensing. 